# NEUTRINO OSCILLATIONS FROM COSMIC SOURCES: A NU WINDOW TO COSMOLOGY

D. J. WAGNER* and THOMAS J. WEILER†

*Department of Physics and Astronomy, Vanderbilt University, Nashville, TN 37235, USA*

Neutrino astrophysics promises a wealth of information about neutrinos and the history of the universe through which they have traveled. In this letter, we extend the standard neutrino oscillation discussion to neutrinos propagating through expanding curved space. This extension introduces a new cosmological parameter in the oscillation phase. The new parameter $\mathcal{I}$ records cosmic history in much the same manner as the redshift variable $z$ or the apparent luminosity distance $D_L$. Measuring $\mathcal{I}$ through neutrino oscillations may help to determine cosmological parameters and distinguish between different cosmologies.

With the advent of neutrino telescopes being constructed around the world, the field of neutrino astrophysics promises unequaled amounts of neutrino data in the coming years. Cosmologically distant sources of neutrinos such as Active Galactic Nuclei (AGNs)[1] and possibly Gamma-Ray Bursters (GRBs)[2] could provide the ultimate long-baseline neutrino experiment. Neutrino oscillations over a cosmic baseline carry an imprint of the universe's history, so neutrino astrophysics offers an entirely new window to cosmology.

Neutrino oscillations will occur if neutrino states of definite flavor, $\nu_\alpha$, are not states of definite mass, $\nu_i$. A neutrino created at time and position $(0,0)$ with a flavor $\alpha$ will be a superposition of all mass states:

$$|\nu(0,0)\rangle = |\nu_\alpha\rangle = \sum_{i=1}^{n} V_{\alpha i}|\nu_i\rangle . \tag{1}$$

Because the propagation of a neutrino depends on its mass, states of different masses will have different energies and momenta and travel at different speeds. The superposition of mass states, and the flavor of the neutrino, will therefore change as the neutrino propagates through space and time. In two generations, the probability that the neutrino has become a different flavor $\beta$ at a time and position $(t,x)$ is[3]

$$P(\nu_\alpha \to \nu_\beta)(t,x) = \sin^2 2\theta \sin^2 \frac{\phi_1 - \phi_2}{2} , \tag{2}$$

*E-mail: wagner@vuhep.phy.vanderbilt.edu
†E-mail: weilertj@macpost.vanderbilt.edu

where

$$\phi_i = E_i t - p_i x \tag{3}$$

is the usual kinematic phase of the $i$th mass state in a static flat-space metric, and $\theta$ is the mixing angle between the two neutrino states $\nu_\alpha$ and $\nu_\beta$.

The relative phase, $\phi_1 - \phi_2$, contains all of the neutrino's kinematic information. The traditional treatment of neutrino oscillations sets $p_1 = p_2 = p$, so the relative phase in static flat space becomes

$$\phi_1 - \phi_2 \approx \frac{\delta m^2}{2p} t, \tag{4}$$

with $\delta m^2 \equiv m_1^2 - m_2^2$. The probability in Eq. (2) thus oscillates in time, with a period of $4\pi p/\delta m^2$ in static flat space. In this letter, we present the correct expression for the relative phase in expanding curved space and examine its significance.

The momentum of a neutrino traveling through curved space redshifts and so is a function of time. The relative phase is therefore given by an integral over the redshift $z$:

$$\phi_1 - \phi_2 = \int_0^{z_e} dz \frac{dt}{dz} \frac{\delta m^2}{2p_0(1+z)} \equiv \frac{\delta m^2}{2p_0} H_0^{-1} \mathcal{I}(z_e). \tag{5}$$

The subscript "$e$" refers to the emission of the neutrino, and the subscript "0" refers to the present. $H_0^{-1} = 3000 h^{-1}$ Mpc is the Hubble time or length, $h \sim \mathcal{O}(1)$ is the Hubble parameter in units of 100 km/s/Mpc, and

$$\mathcal{I}(z_e) = H_0 \int_0^{z_e} \frac{dt}{dz} \frac{dz}{1+z} = H_0 \int_1^{1+z_e} \frac{d\omega}{\omega^2} \frac{1}{H(\omega)} \tag{6}$$

may be thought of as the fraction of the Hubble time $H_0^{-1}$ available for neutrino transit and oscillation. $H(\omega)$ is the Hubble parameter at cosmic redshift $z = \omega - 1$. Comparing Eqs. (4) and (5), we see that the correct generalization of the oscillation probability from Minkowski space to an expanding cosmology is obtained by replacing the neutrino's transit time with the expansion-weighted time $H_0^{-1}\mathcal{I}$. We have also derived this general-relativistically correct expression for the relative phase in two other ways: integrating the covariant action along the geodesic of the Robertson–Walker metric, and solving the equation of motion of a scalar field for the phase in curved space–time. Both re-derivations agree with the simple result (5) to the order of interest, $\delta m^2/p^2$.[4]

Equation (6) implies a bound of $H_0^{-1}\mathcal{I}(z_e = \infty)$ for the maximum "distance" available for neutrino oscillations in our universe. We will show below that a significant fraction of this ultimate baseline is already encountered by neutrinos emitted by objects at redshifts of 1 or 2, just the redshifts attributed to GRBs and distant AGNs.

The relative phase in Eq. (5) is no longer linear in time, so the oscillation is no longer uniform. Oscillation periods are given by the differences between two zeros of the oscillatory term. Setting

$$\frac{\delta m^2 H_0^{-1}}{4p}(\mathcal{I}(z_e) - \mathcal{I}(z_n)) = n\pi \tag{7}$$

yields the locations $z_n$ of the nodes of the oscillation. Since the location of the nodes is energy-dependent, the Earth sits at an oscillation node for some particular energy. At that energy we will observe the neutrino flavors in just the ratios with which they were produced at the cosmic source.

The factor $\mathcal{I}$ presents us with a new cosmological variable. It records cosmic history in much the same manner as the redshift variable $z$, or the apparent luminosity distance $D_L$. For small $z$, we may make a Taylor expansion of Eq. (6) around $z = 0$:

$$\mathcal{I}(z) = z - \frac{1}{2}(3 + q_0)z^2 + \cdots \tag{8}$$

and

$$z(\mathcal{I}) = \mathcal{I} + \frac{1}{2}(3 + q_0)\mathcal{I}^2 + \cdots , \tag{9}$$

where $q_0$ is the standard deceleration parameter, describing the slowing of expansion. Similar expressions may be found for the luminosity distance $D_L$ in terms of $z$ or $\mathcal{I}$. To first order, the variables $z$, $\mathcal{I}$, and $D_L$ are linearly related; to first nonlinear order, the relations among the three variables are determined completely by $q_0$, a currently unmeasured quantity. The value of $q_0$ is actively being sought with automated telescopes using Type 1A supernovae as standard candles.

Studying cosmology by measuring neutrino flavor ratios has one tremendous advantage over other methods, since flavor ratios should be independent of any evolutionary effects in the ensemble of sources. Deviations from the linear Hubble law inferred through the use of distant candle luminosities may exhibit evolutionary effects. Initial neutrino flavor ratios, however, are fixed by microphysics and should not change with cosmic history, even if the absolute neutrino luminosities evolve. Oscillation measurements with AGN or GRB neutrinos may therefore shed some "neutrino light" on the important parameter $q_0$.

For $z \gtrsim 0.5$, the oscillation phase is sensitive to the cosmological model. Assuming a standard hot big-bang Friedmann cosmology,

$$H(\omega) = H_0 \left[\Omega_r \omega^4 + \Omega_m \omega^3 - k\Omega_k \omega^2 + \Omega_\Lambda\right]^{1/2} \tag{10}$$

and

$$\Omega_r + \Omega_m - k\Omega_k + \Omega_\lambda = 1. \tag{11}$$

$\Omega_{r/m}$ are the present radiation and matter densities in units of the critical density. $\Omega_k = (R_0 H_0)^{-2}$ is the curvature term, with $k = 1, 0, -1$ for a closed, flat, or open universe, and $\Omega_\Lambda = \Lambda/(3H_0^2)$ is the scaled cosmological constant. Our universe is not radiation-dominated, so we may neglect the $\Omega_r$ term. Inflation favors cosmologies for which $k = 0$; we will here consider only these zero-curvature possibilities.

In a Friedmann cosmology with zero curvature and radiation,

$$\mathcal{I}(z_e) = H_0^{-1} \int_1^{1+z_e} \frac{d\omega}{\omega^2} \left[\Omega_m \omega^3 + \Omega_\Lambda\right]^{-1/2}, \tag{12}$$

with $\Omega_m + \Omega_\Lambda = 1$. The special case with zero cosmological constant is easily solved analytically:

$$\mathcal{I}(z_e) = \frac{2}{5} H_0^{-1} \left[1 - (1+z_e)^{-5/2}\right]. \tag{13}$$

In this case, $\mathcal{I}$ asymptotically approaches $\frac{2}{5} H_0^{-1}$ as $z$ increases. For neutrinos emitted from an object with redshift of $z = 1$, $\mathcal{I}$ is already $0.33 H_0^{-1}$, or 83% of its asymptotic value.

Examples of $\mathcal{I}(z_e)$ versus $z_e$ for different values of $\Omega_m$ and $\Omega_\Lambda = 1 - \Omega_m$ are shown in Fig. 1. This figure demonstrates the asymptotic behavior of $\mathcal{I}$ for all values of $\Omega_m$ and illustrates that $\mathcal{I}$ has already attained a significant fraction of its asymptotic value at redshifts attributed to GRBs and distant AGNs, $z \gtrsim 1$.

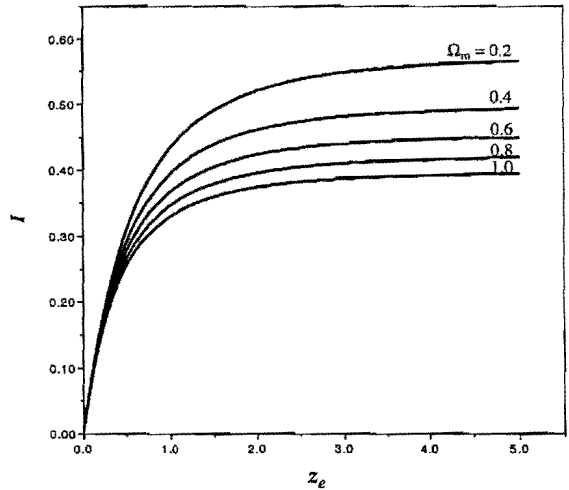

Fig. 1. $\mathcal{I}$ as a function of $z_e$ for different zero-curvature cosmologies. The cosmological constant may be found from $\Omega_\Lambda = 1 - \Omega_m$.

The asymptotic behavior at the redshifts typical of GRBs and AGNs has an important implication for neutrino physics: since the value of $\mathcal{I}$ in the oscillation phase is a known fraction of $H_0^{-1}$ for a given cosmology, the uncertainty in $\mathcal{I}$ is dominated by the uncertainty in $H_0$, which is less than a factor of two and certain to decrease in the near future. Thus, a deduction of the oscillation phase and neutrino energy may permit a determination of $\delta m^2$'s to better than a factor of two! This cosmic baseline oscillation measurement is the only way to probe the tiny $\delta m^2$'s predicted by some particle physics models.

Comparing an exact cosmological solution for $z$, $\mathcal{I}$, or $D_L$ to the appropriate Taylor series approximation relates each term in the series to more fundamental quantities. The simplest relation is between the parameter $q_0$ and the parameters $\Omega_m$ and $\Omega_\Lambda$ in the Friedmann equation: $q_0 = \Omega_m/2 - \Omega_\Lambda$. Thus, a neutrino oscillation determination of $q_0$ would establish a fundamental constraint between the two parameters of the Friedmann universe.

Neutrino oscillations promise a new avenue to explore the universe. Just as measuring a photon energy can yield the redshift of the source, measuring the flavor ratios and energies of neutrinos at the earth may yield the quantity $\mathcal{I}$. The former is possible if the energy of the photon at the source is known; for the latter to work, the flavor ratio at emission and other mixing parameters must be known. Neutrino oscillation measurements are possible at proposed neutrino telescopes.[5] These measurements, combined with independent measurements of $z$ or $D_L$ will potentially produce a strong constraint on the cosmological model and determine both $\delta m^2$ and $q_0$.

## Acknowledgments

This work was supported in part by the U.S. Department of Energy grant No. DE-FG05-85ER40226 and by the NASA Tennessee Space Grant Consortium.